\documentclass{PoS}
\usepackage{wrapfig}
\usepackage{subfig}

\title{Two-Color Schr\"{o}dinger Functional with Six-Flavors of Stout-Smeared Wilson Fermions}

\ShortTitle{Two-Color Schr\"{o}dinger Functional with Six-Flavors of Stout-Smeared Wilson Fermions}

\author{\speaker{Gennady Voronov} (for the LSD Collaboration) \\%
        Department of Physics, Yale University, New Haven, CT 06511, USA\\
        E-mail: \email{gennady.voronov@yale.edu}}

\abstract{
We study the Schr\"odinger functional running coupling in the $SU\left(2\right)$
gauge theory with six-flavors of massless fermions. The aim is to
determine whether the above theory has an infrared fixed point (IRFP).
We use the standard Wilson gauge action and the stout-smeared Wilson
fermion action. Here we present a determination of the critical mass
as a function of the bare coupling and a preliminary study of the
phase diagram of this lattice action. We also find preliminary indication
that this theory has no IRFP. While this conclusion is not yet definite,
we also show that with this approach we will be able to take a proper
continuum limit and clearly determine the status of this theory with
a reasonable amount of computer time.
}

\FullConference{The 30th International Symposium on Lattice Field Theory\\
                 June 24 – 29,  2012\\
                 Cairns, Australia}

\begin{document}

\section{Introduction }

A SM Higgs adequately accounts for all electroweak measurements (for
now). However, it has a number of theoretical shortcomings, chiefly
among them the hierarchy problem. technicolor is an alternative for
electroweak symmetry breaking that avoids the introduction of a fundamental
light scalar particle \cite{hill01}. In particular, a technicolor
model based on a walking gauge theory can perhaps account for the
standard model fermion masses \cite{appelquist01}. Since such a theory
is expected to reside just below the conformal window, it is essential
to narrow down the extent of this window. The goal of this work is
to do that for the $SU(2)$ gauge theories with $N_{f}$ fermion flavors
in the fundamental representation.

The $SU\left(2\right)$ gauge theories are particularly interesting
in light of the recent discovery of a 125 GeV Higgs-like particle
\cite{atlas01,cms01}. This newly discovered resonance , while not
yet confirmed, is widely expected to have the quantum numbers of a
scalar particle. In the framework of strongly coupled theories, light
scalars arise as pseudo-Nambu-Goldstone bosons (PNGB), i.e. from explicit
breaking of a continuous symmetry which would be broken spontaneously
in the absence of explicit breaking. One speculative mechanism is
that the breaking of approximate scale invariance, as is present in
a walking gauge theory will produce a light dilaton \cite{appelquist02}.
This gives yet another motivation to search for walking theories.
PNGBs are also expected from the breaking of global chiral symmetries.
In QCD, this mechanism yields light pseudo-scalars. Two-color theories,
due to the pseudo reality of the fundamental representation of $SU\left(2\right)$,
have an enhanced chiral symmetry and hence novel symmetry breaking
patterns \cite{Peskin01}. A number of authors use this novel breaking
pattern to construct strongly coupled models whose spectrum contain
several light scalar PNGBs \cite{luty01,Katz01}, the lightest of
which would be a suitable Higgs candidate. 

We take a rather direct route towards narrowing down the extent of
the conformal window for two-color theories with $N_{f}$ flavors
of Dirac fermions. We calculate the running coupling in the Schr\"{o}dinger
functional ($SF$) scheme \cite{Luscher01} and directly search for
the presence or lack of an infrared fixed point (IRFP) at various
$N_{f}$. Evidence that $N_{f}=10(4)$ is inside (outside) the conformal
window is presented by \cite{karavitra01}. Additionally Ohki et.
al. argue that $N_{f}=8$ is inside the conformal window \cite{Ohki01}.
The case $N_{f}=6$ while tackled by many groups \cite{karavitra01,Bursa01,Voronov01}
remains inconclusive. 

In \cite{Voronov01}, we attempt to determine whether $N_{f}=6$ has
an IRFP. We used the standard Wilson plaquette gauge action and the
Wilson fermion action. Our calculation was inconclusive due to mistuning
of the bare fermion mass and an inability to probe sufficiently strong
renormalized couplings at computationally feasible lattice volumes
due to a lattice artifact driven phase transition. Here we revisit
the $N_{f}=6$ theory but we move to the stout-smeared \cite{morn01}
Wilson fermion action. This action avoids coupling the fermions to
the unphysical fluctuations of the gauge field on the scale of the
lattice spacing. This approach allow us to probe a sufficiently large
renormalized couplings to either see an IRFP or definitely rule one
out. Additionally, we undertake a great deal more careful tuning of
the bare fermion mass.

\section{Preliminary Study of the Lattice Action}

For this work we use the standard Wilson plaquette gauge action and
the stout-smeared Wilson fermion action. The Wilson fermion action
contains an additional irrelevant operator that lifts the mass of
the fermion doublers to the cutoff scale so they decouple from the
calculation. This additional term explicitly breaks chiral symmetry
and as a result the fermion mass is additively renormalized. As a
result the bare mass must be carefully tuned in order to restore chiral
symmetry. This critical value of the bare mass (as a function of the
bare coupling) $m_{c}(g_{0}^{2})$ is defined as the bare mass value,
$m_{0}$, that results in a zero PCAC quark mass \cite{luscher02}.
In practice, $m_{c}$ is determined, at fixed bare gauge coupling
$g_{0}^{2}$ and lattice volume $\left(L/a\right)^{3}\times2L/a$,
as the root of a fitted linear function to measurements of the PCAC
quark mass versus the bare quark mass. One example of this procedure
is shown below in figure 1. This is done for a range of bare couplings
and lattice volumes and fit to a polynomial given by 
\begin{equation}
m_{c}^{\mbox{fit}}\left(g_{0}^{2},\frac{a}{L}\right)=\sum_{i=1}^{n}g_{0}^{2i}\left[a_{i}+b_{i}\left(\frac{a}{L}\right)\right].\label{eq:mcfitpoly}
\end{equation}
 Finally, $m_{c}^{\mbox{fit}}\left(g_{0}^{2},0\right)$ will be used
in all following running coupling calculations. All data used to fit
$m_{c}^{\mbox{fit}}$ and $m_{c}^{\mbox{fit}}\left(g_{0}^{2},0\right)$
show below in figure 2. 
\begin{figure}
\begin{centering}
\includegraphics[width=8cm]{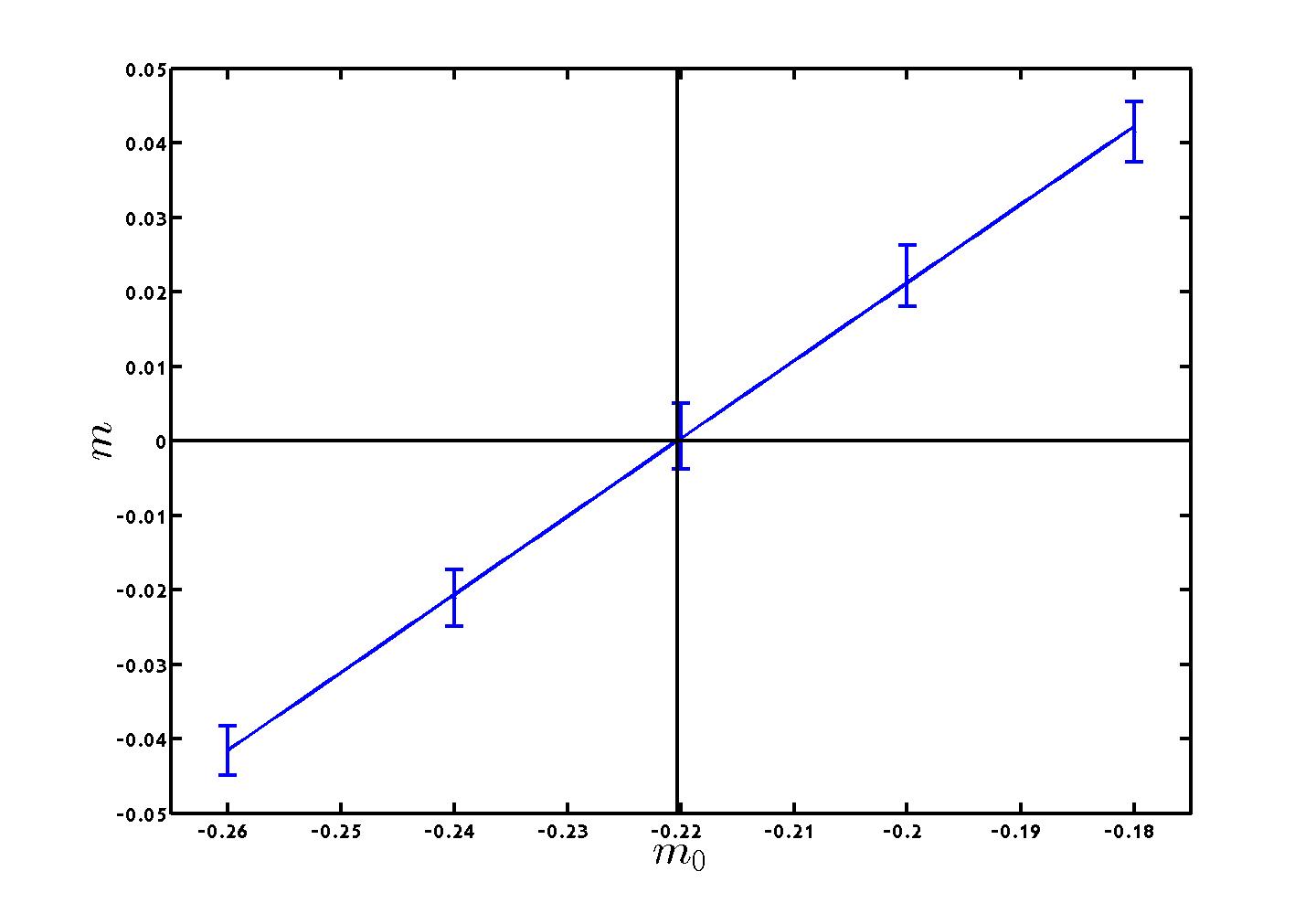}
\par\end{centering}

\caption{PCAC quark mass at $g_{0}^{2}=1.4$ and $L/a=12$ plotted versus the
bare quark mass $m_{0}$. A linear fit to these points is included
as well. The root of this linear function determines the critical
mass $m_{c}\left(g_{0}^{2}=1.2,\frac{L}{a}=12\right)$, the value
of the bare mass which results in restored chiral symmetry signaled
by a zero PCAC mass. The root is highlighted in this figure by a vertical
line. This figure is representative of all critical mass determinations
at different bare couplings and lattice volumes. }
\end{figure}
 
\begin{figure}
\begin{centering}
\includegraphics[width=8cm]{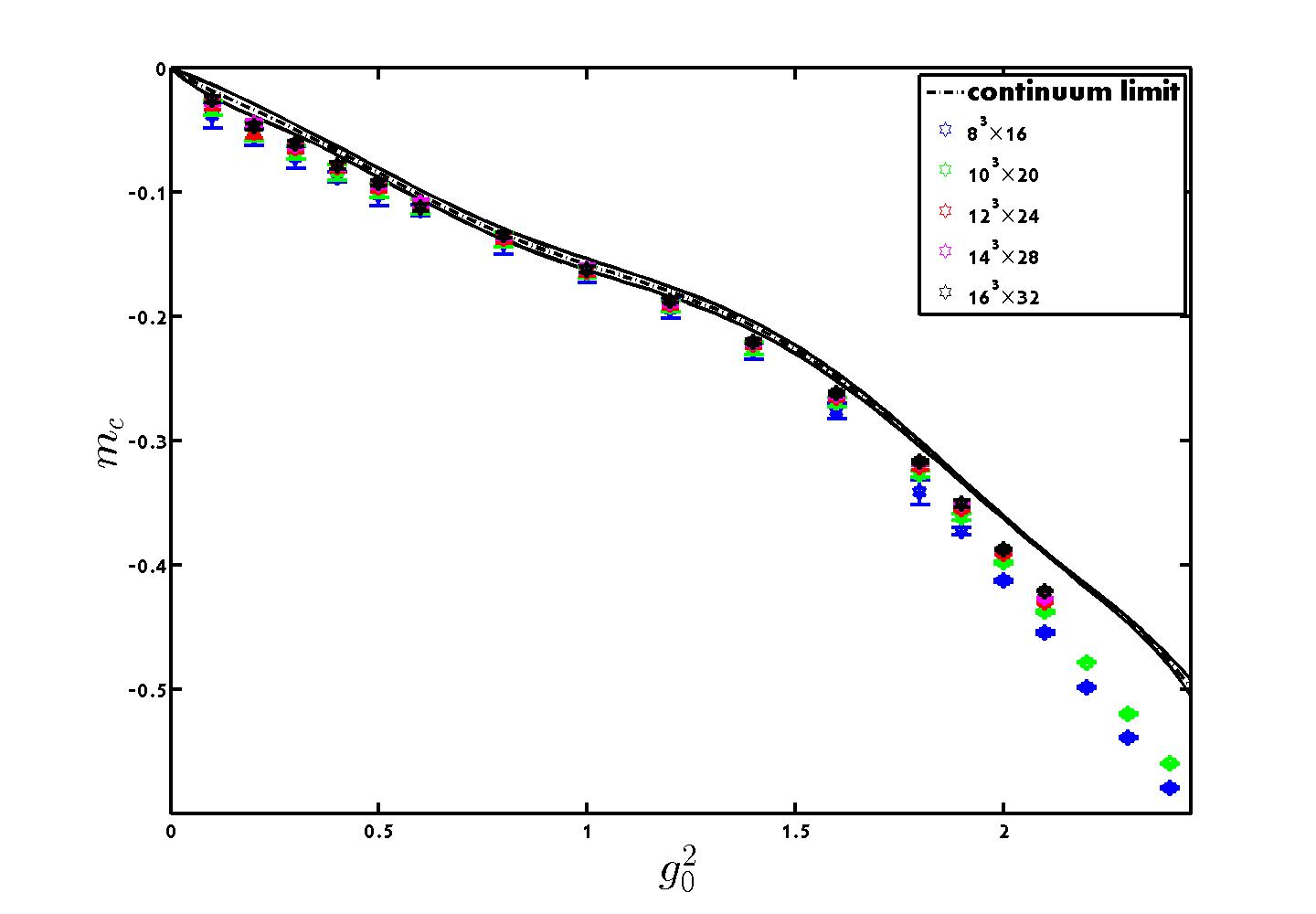}\caption{All critical masses determined by similar procedure to that which
is shown in figure 1. All $m_{c}$ determinations are fit to the polynomial
$m_{c}^{\mbox{fit}}\left(g_{0}^{2},\frac{a}{L}\right)$ given in eq.
(2.1). We show an extrapolation to the continuum critical mass which
is given by $m_{c}^{\mbox{fit}}\left(g_{0}^{2},0\right)$. }

\par\end{centering}

\end{figure}

In order to guarantee that we can take a continuum limit, we need
to ensure that we obtain data from the weak-coupling side of any spurious
lattice phase transition. With this in mind, we scanned through the
bare parameter space and located peaks in the plaquette susceptibility
on a $L/a=10^{3}\times11$ lattice. This search indicates a line in
the $m_{0}-g_{0}^{2}$ plane of first order phase transition that
ends at a critical point at around $g_{0}^{2}\approx2.2$. For $g_{0}^{2}\lesssim2.3$,
we see crossover behavior. In Figure 3, we show the above transition
line plotted along with $m_{c}^{\mbox{fit}}(g_{0}^{2},0)$. Figure
3 indicates that the six flavor massless fundamental Wilson fermion
action has a sensible continuum limit only for $g_{0}^{2}\lesssim2.175$.
Therefore, we will only examine the running coupling on lattices with
a bare coupling $g_{0}^{2}<2.175$. 
\begin{figure}
\begin{centering}
\includegraphics[width=8cm]{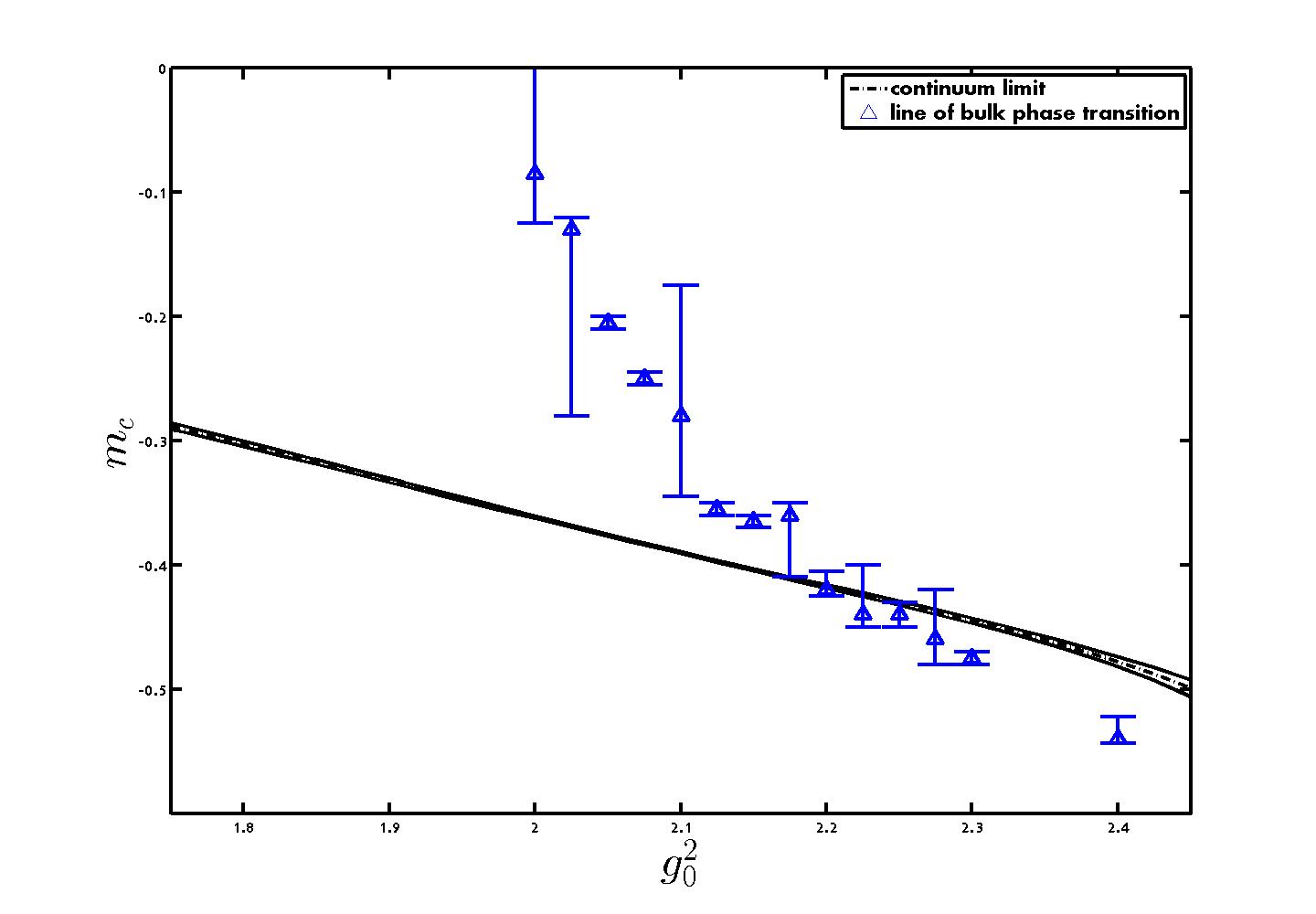}\caption{$m_{c}^{\mbox{fit}}\left(g_{0}^{2},0\right)$ plotted along with the
peak in the plaquette susceptibility. We collect all running coupling
data along the critical mass line on the weak coupling side of the
phase transition line. }

\par\end{centering}

\end{figure}

The final choices required to specify our action are detailing the
gauge field smearing operator used in the fermion action. We use only
one level of stout-smearing with a smearing parameter $\rho=0.25$.
These properties of the smearing operator are described in \cite{morn01}.
As all calculations in this work are done with Dirichlet boundary
conditions in the time directions, there is some ambiguity in how
to implement the smearing of the gauge field near this boundary. It
is clearly important to not smear boundary links with bulk links.
Even with this constraint there is still some ambiguity in how we
define our smearing operator. Specifically we can choose whether to
allow the bulk links near the boundary to be smeared with boundary
links. 

The chief observable of interest is the $SF$ running coupling, roughly
it is the derivative of the action with respect to the $SF$ boundary
conditions \cite{luscher03}. If we smear bulk links with boundary
links, then this observable becomes significantly more complicated
and difficult to calculate. Hence we would prefer to define our smearing
operator to avoid this. We did not fully appreciate this issue until
after completing the preliminary study of the phase diagram of our
action. However we recomputed the critical mass and the plaquette
susceptibility on a limited sample with the preferable definition
of the smearing operator. We show a comparison between the critical
mass calculated with the two different smearing procedures in figure
4. Additionally, no significant shift in the phase boundary shown
in figure 3 was found. A comparison of the peak susceptibility (used
in determining the phase boundary) on a sample of the data is shown
in figure 5. We conclude that any shift in quantities used to determine
the phase diagram, due to modifying the smearing near the boundaries,
is well below all other sources of errors and we proceed to use the
preferable smearing definition for the running coupling while using
the determined calculated critical mass and phase boundary. 
\begin{figure}
\begin{centering}
\subfloat[]{\begin{centering}
\includegraphics[width=8cm]{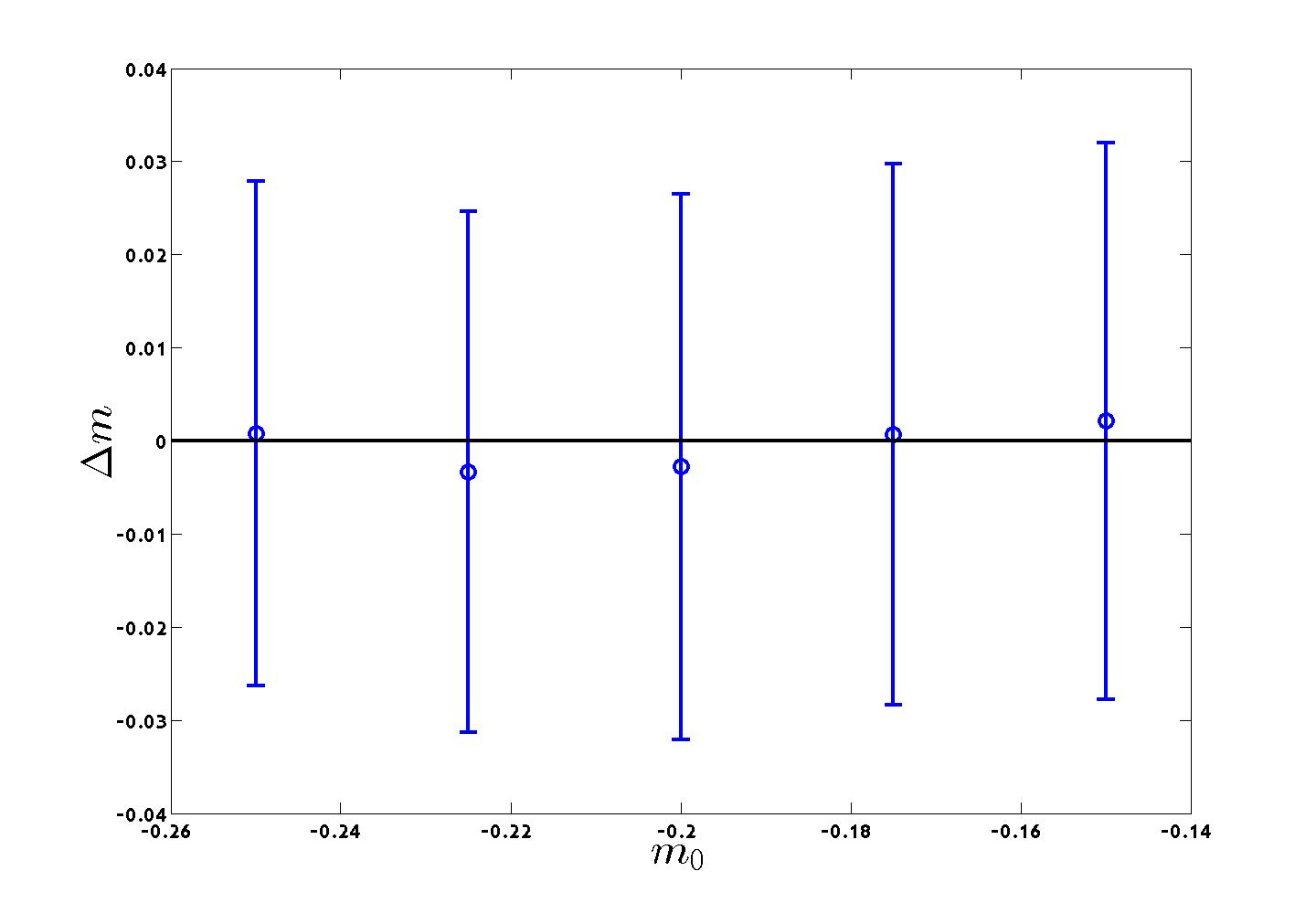}
\par\end{centering}

\centering{}} \subfloat[]{\begin{centering}
\includegraphics[width=8cm]{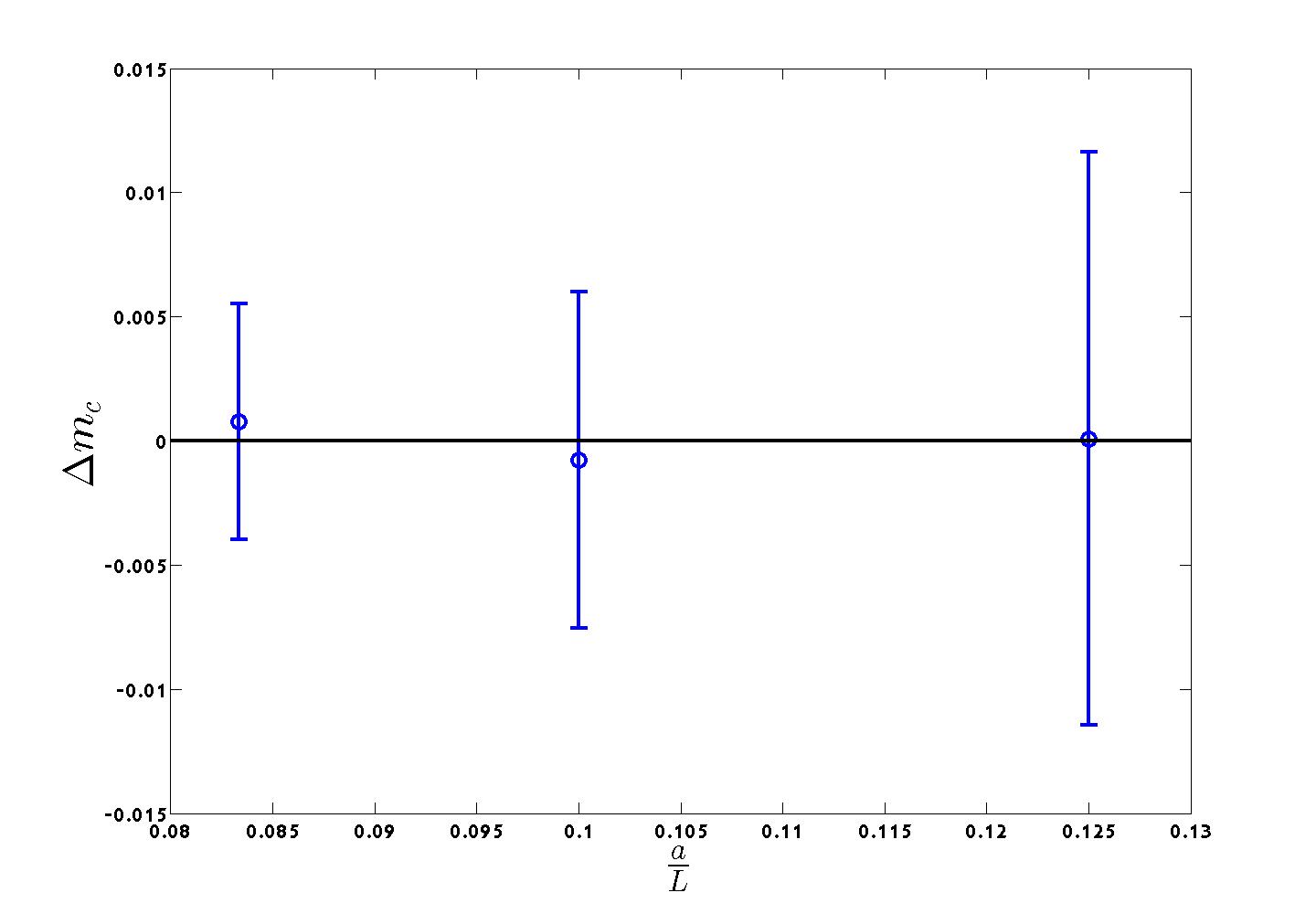}
\par\end{centering}

}
\par\end{centering}

\caption{A representative sampling of differences, when calculated using two
different definitions of the smearing operator near the boundaries
in the time directions, in two quantities relevant in the analysis
of the critical mass. All data shown in (a) and (b) obtained at $g_{0}^{2}=1.2$
In (a) we show the difference in the PCAC quark mass $m$ at $L/a=8$
versus several bare quark masses $m_{0}$ near the critical mass.
In (b), the difference in the critical mass $m_{c}\left(g_{0}^{2}=1.2,\frac{L}{a}\right)$
is shown versus $L/a$. Both (a) and (b) show no statistically significant
difference.}
\end{figure}
 
\begin{figure}
\begin{centering}
\includegraphics[width=8cm]{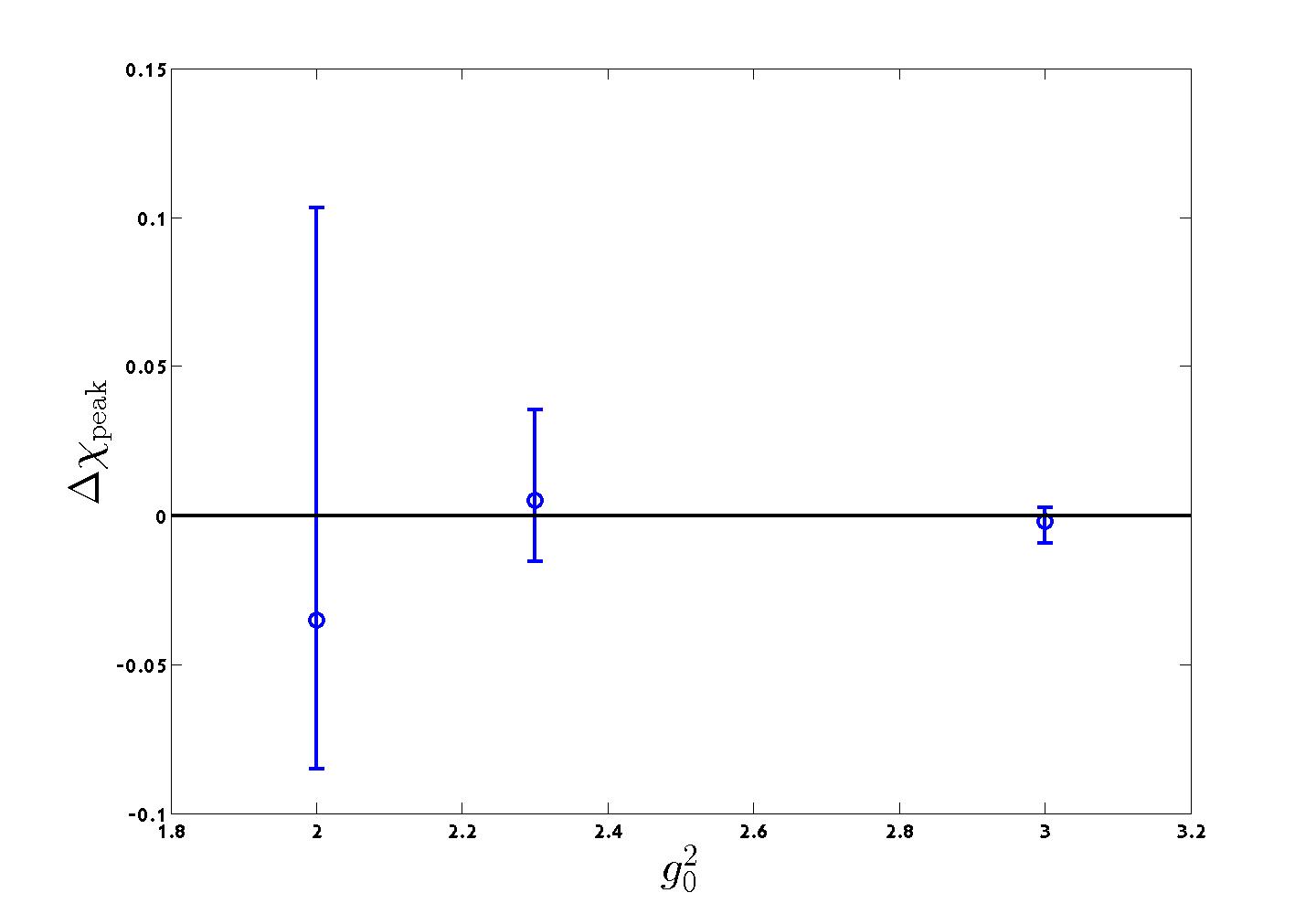}
\par\end{centering}

\caption{Difference in determination of the peak plaquette susceptibility,
when calculated using two different definitions of the smearing operator
near the boundaries in the time directions, versus $g_{0}^{2}$. All
points calculated on lattice volumes $L/a=10^{3}\times11$. No statistically
significant differences found.}
\end{figure}

\section{Running Coupling and Conclusions}

We calculate the $SF$ running coupling, as defined here \cite{luscher03},
at several fixed bare coupling and a variety of lattice volumes. Results
are show in figure 6. No attempt is made to take a continuum limit
hence no rigorous conclusions about the IR nature of this theory can
be made. However this plot does give a preliminary indication that
the two-color six-flavor gauge theory is confined and chirally broken
in the IR. This can be inferred from the lack of evidence of an IRFP
up to renormalized couplings well beyond the strength that would be
required to break chiral symmetry dynamically \cite{Cohen01}. Moreover
the renormalized coupling crosses this threshold on lattice volumes
as low as $L/a=10$ in the range of bare couplings we are free to
probe. This implies that an accurate determination of the IR status
of this theory is computationally feasible with our choice of action.
The qualitative behavior of figure 6 stands in contrast to that of
figure 1 of \cite{hietanen01}. There a theory, which is widely believed
to be inside the conformal window and whose renormalized couplings,
when plotted in the manner of figure 6 here, show behavior that is
consistent with the existence of an IRFP. 
\begin{figure}
\begin{centering}
\includegraphics[width=8cm]{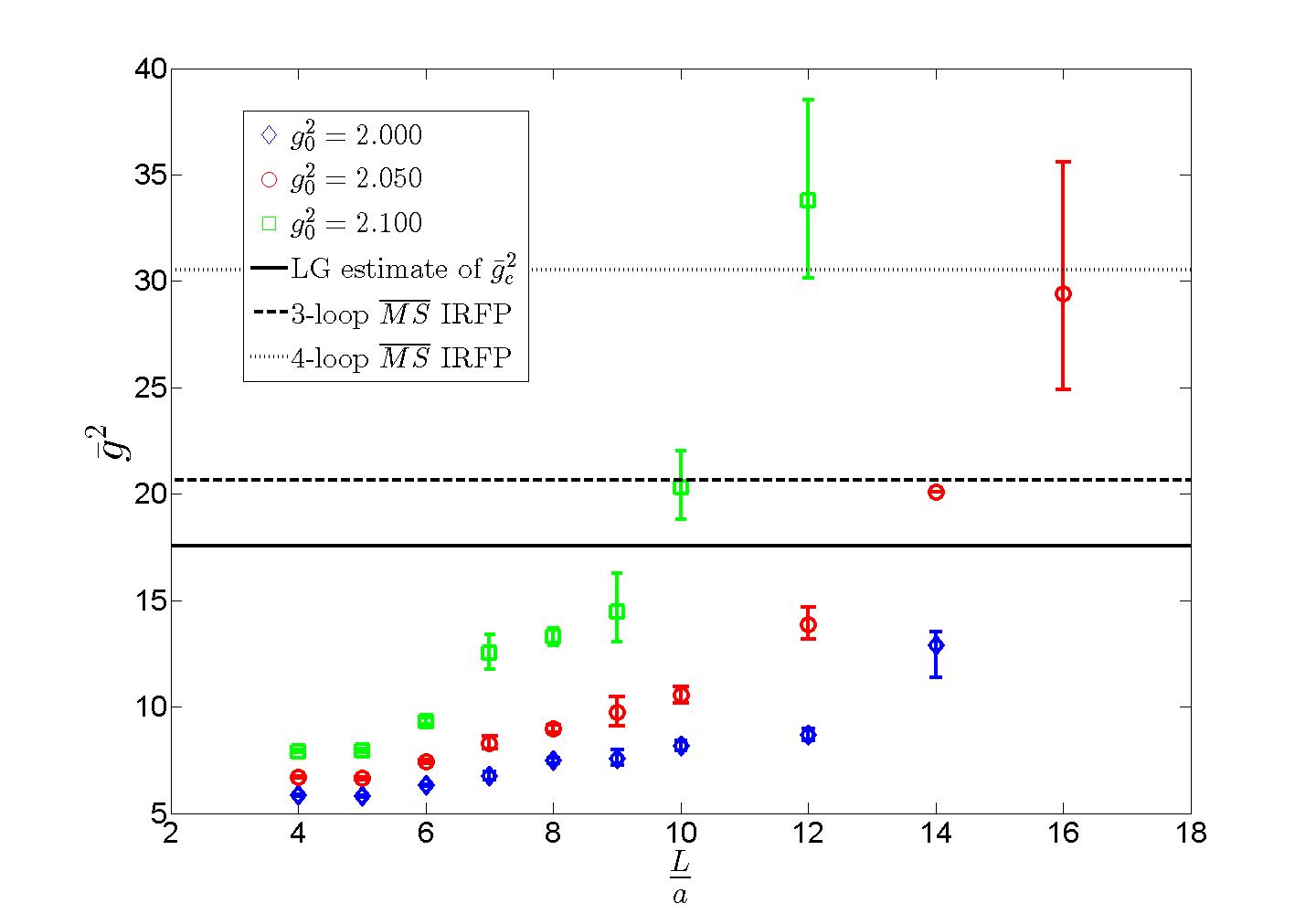}\caption{The $SF$ running coupling evaluated, at fixed $g_{0}^{2}=2.0,$ $2.05$,
and $2.1$, versus $L/a$. The ladder gap estimate of the critical
renormalized coupling strength required to break chiral symmetry is
shown along with the 3 and 4 loop $\overline{MS}$ scheme IRFP strengths.
The 2 loop IRFP strength is $\bar{g}_{*}^{2}\approx144$.}

\par\end{centering}

\end{figure}

Of course an IRFP can appear once we take a continuum limit. To obtain
such a limit, we would need to produce a series of curves like those
in figure 6 but at many more values of the bare coupling. The finally
step would be to do a step scaling analysis as in \cite{lusher04,Voronov01}.
This is currently in progress. 

We have presented on the current status of our $SF$ running coupling
calculation of the two-color six-flavor gauge theory. A thorough study
of the parameter space of our action has been undertaken and we are
confident that we have sufficiently tuned the bare fermion mass. The
phase diagram has also been studied to ensure that all results are
inferred from the weak side of any spurious lattice phase transition.
Study of the phase diagram in conjunction with our preliminary running
coupling calculation imply that the definite determination status
of this theory is computationally feasible. Moreover, we see a preliminary
indication that this theory has no IRFP, but a definite conclusion
will have to wait.

\end{document}